\documentclass[aps,prb,twocolumn,groupedaddress,showpacs]{revtex4}

\usepackage{graphicx}
\usepackage{dcolumn}
\usepackage{bm}
\usepackage{amsmath}

\begin{document}


\title{Self-organized current transport through low angle grain boundaries in 
YBa$_2$Cu$_3$O$_{7-\delta}$ thin films, studied magnetometrically}



\author{J.R. Thompson}
\affiliation{Condensed Matter Sciences Division, Oak Ridge National Laboratory, Oak Ridge, Tennesee 37831-6061 and Department of Physics, The University of Tennessee, Knoxville, Tennessee 37996-1200}

\author{H.J. Kim}
\altaffiliation{now at Department of Physics and Astronomy, 
Michigan State University, East Lansing, Michigan}
\affiliation{Department of Physics, The University of Tennessee, Knoxville, Tennessee 37996-1200}

\author{C. Cantoni, D.K. Christen, R. Feenstra}
\author{D.T. Verebelyi}
\altaffiliation{now at American Superconductor Corp., Westborough, Massachusetts}
\affiliation{Condensed Matter Sciences Division, Oak Ridge National Laboratory, Oak Ridge, Tennesee 37831-6061 }




\date{15 August 2003}

\begin{abstract}
The critical current density flowing across low angle grain boundaries in YBa$_2$Cu$_3$O$_{7-\delta}$ thin films has been studied magnetometrically.  Films (200 nm thickness) were deposited on SrTiO$_3$ bicrystal substrates containing a single [001] tilt boundary, with angles of 2, 3, 5, and 7 degrees, and the films were patterned into rings.  Their magnetic moments were measured in applied magnetic fields up to 30 kOe at temperatures of 5 - 95 K; current densities of rings with or without grain boundaries were obtained from a modified critical state model.  For rings containing 5 and 7 degree boundaries, the magnetic response depends strongly on the field history, which arises in large part from self-field effects acting on the grain boundary.
\end{abstract}


\maketitle

\section{INTRODUCTION}
High temperature superconductors characteristically have highly anisotropic properties and a short scale of the superconducting coherence length $\xi$.  As a consequence, the critical current density across grain boundaries can be significantly reduced relative to that which flows within grains.  For reasons that are not well understood, a large misalignment of adjacent grains suppresses the order parameter along the grain boundary (GB) and as a result, the adjacent grains are weakly linked.  The weak-link behavior of a high-angle grain boundary and a near-exponential decrease in the GB critical current density $J_c^{GB}$ with misorientation angle were first studied by Dimos et al. \cite{Dimos88} in YBa$_2$Cu$_3$O$_{7-\delta}$ materials.  Since then, a number of studies have been conducted on this and other high-$T_c$ superconductors, as reviewed recently by Hilgenkamp and Mannhart.\cite{HM-RMP02}  This review article provides an excellent overview of grain boundaries in these materials, the controlled synthesis of GBs, current transport, and related topics.  Much earlier work focussed on the properties of GBs with relatively large misorientation angles $\theta > $ 10$^{\circ}$, where intergrain current conduction is severely suppressed. The present work concentrates on materials with lower angle GBs, which are contained in YBCO rings for magnetometric study.

The regime of lower angle grain boundaries is interesting, for with decreasing tilt angle $\theta$, the mode of current conduction appears to cross-over from weak to strong linkage.  Technologically this is highly relevant: recently developed methods for forming highly-textured coated conductors, such as rolling-assisted biaxially textured substrates (RABiTS)\cite{Goyal96,Goyal99} and ion-beam assisted deposition (IBAD)\cite{Iijima92,Reade93,Wu95}, vastly improve the current conduction in multicrystalline coated conductors of high-temperature superconductors, by reducing the average misorientation into a regime of low angles $\theta  \lesssim 7^{\circ}$.  For sufficiently small angles, the material between dislocation cores on the grain boundary is only mildly perturbed and this provides a strong conduction channel comparable to or wider than the in-plane coherence length.\cite{Verebelyi00}  For further development of coated conductors, an understanding of low-angle GBs is important, for it gives guidance as to how highly textured the materials must be.  Specifically, are low-angle GBs still weak-linked and responsible for large reductions of the GB current density?  How does the application of an external magnetic field affect the current flow? 

This magnetometric study of low-angle GBs is based on a simple equation from electrodynamics.  A circulating current generates a magnetic dipole moment according to the equation
\begin{equation}
m = (1/2c) \int (\boldsymbol{r} \times
          \boldsymbol{J} (\boldsymbol{r}))  \text{d}V	
\label{eq:one}
\end{equation}
where $\boldsymbol{J}$ is the current density at location $\boldsymbol{r}$. This expression tells us that once the current configuration is established, the current or (spatially uniform) current density can be calculated from the measured magnetic moment of a sample.  The current configuration is related to the sample geometry.  With the sensitivity of SQUID-based instrumentation and a simple ring geometry that helps define the current path,\cite{Jung93,Darhmaoui96,Charalambous98,Claus01} we obtain the critical current density $J_c$ flowing through grain boundaries in the high-temperature superconductor YBa$_2$Cu$_3$O$_{7-\delta}$ (YBCO), and for comparison, the $J_c$ of companion rings with no GBs. Furthermore, the magnetic responses $m(H)$ of the rings are compared and contrasted.  We observe a large peak in the decreasing field branch of the $m(H)$ curves for GB rings with $\theta$ = 5.1 and 7$^{\circ}$ and show that its appearance arises largely from a cancellation of the applied magnetic field by self field effects on weakly linked grain boundaries.  

\section{MAGNETIC MOMENTS in the CRITICAL STATE}

In the critical state model, one assumes that critical currents with density $\pm J_c$ flow throughout the superconductor.  In this work, the currents are induced by applying a large magnetic field $H_{app}$ perpendicular to flat, planar samples.  It is then straightforward to integrate Eq. 1 and obtain the associated magnetic moment $m$ for several of the sample geometries used here.  For example, a flat strip of thickness $d$, length $\ell$, and width $w$ has 
\begin{equation}
m_{strip}=J_c d \ell w^2 /40 \times [1-w/3 \ell ]   		
\label{eq:strip}
\end{equation}
according to the sandpile model.\cite{Gyorgy}  This and following expressions have cgs units with dimensions in cm, $J$ in A/cm$^2$, and $m$ in erg/G = emu.  For the long narrow strips considered here, the factor $[1-w/3 \ell ] \approx 1$ will be neglected.  Equation \ref{eq:strip} also gives the moment of an ``open circuit'' thin ring of the same thickness and width, where $\ell = 2 \pi R$ for a ring of radius $R$. 
\begin{equation}
m_{open~ring}=J_c d (2 \pi R) w^2 /40				
\label{eq:openring}
\end{equation}
Another standard case is that of a disk of outer radius $a$ and thickness $d$.  Here one has 
\begin{equation}
m_{disk} = J_c \pi d a^3 /30
\end{equation}
From this, it follows that a continuous ring with outer radius $a$ and inner radius $a_1$ has moment
\begin{equation}
m_{ring} = J_c \pi d (a^3 - a_{1}^{3}) /30		
\label{eq:ring}
\end{equation}
For a thin ring with $a-a_1 =  w << a$, Eq.~\ref{eq:ring} is nearly the same as the simplest expression
\begin{equation}
m_{loop} = I_c \pi a^2/c 						
\label{eq:loop}
\end{equation}
where $I_c$ is the critical current and the speed of light $c$ has value 10 in these laboratory units.

Let us now consider the case of a thin ring that crosses a grain boundary (GB).  The GB is expected to have a lower critical current density $J_{c}^{GB}$ and lower critical current $I_{c}^{GB}$ than the surrounding epitaxial YBCO grain film with critical current density $J_c^{Gr}$. When applying a magnetic field, we first induce currents that circulate around the outer and inner circumferences of the ring and that screen flux from the central hole;\cite{Brandt97} simultaneously flux penetrates into the grain YBCO and more deeply into the GB.  When the current exceeds $I_{c}^{GB}$, flux enters the hole along the GB and is trapped there.  Application of still higher fields drives flux into all of the grain material where currents with density $J_c^{Gr}$ flow throughout.  Thus a portion of the current approaching a GB can cross it and generate the magnetic moment of a loop, Eq.~\ref{eq:loop}.  The self-organizing, remaining portion of the current makes a ``U-turn,''\cite{Polyanskii96} giving the magnetic moment of a strip, Eq.~\ref{eq:strip}.  The resulting Bean-like flux profile is displaced from the center of the strip, just like the case of a superconducting strip carrying both a transport current and critical state currents.  Let us define $\Delta$ by $J_c^{GB} = (\Delta /w)  J_c^{Gr}$ as a measure of the GB current (geometrically $\Delta /2$ corresponds to the displacement of the flux profile from the center-line of the 'strip').
Then we have 
\begin{equation}
m_{GB~ring} = m_{loop} + m_{strip}
\end{equation}
\begin{equation}
m_{GB~ring} =  (\Delta /w)  J_c^{Gr}wd \pi a^2/c + J_c^{Gr} d (2 \pi a) (w- \Delta )^2 /40
\label{eq:GBring}
\end{equation}
These expressions ignore terms of order $(w/a)^2$.  Experimentally, we determine $ J_c^{Gr}$ in a separate experiment, then solve Eq.~\ref{eq:GBring} for $(\Delta /w)$, from which $J_c^{GB}$ is obtained.

\section{EXPERIMENTAL ASPECTS}
	
In order to use the above equations effectively for this study, samples were specially designed as rings in the following process.  Films of YBCO were prepared on SrTiO$_3$ (STO) [001]-tilt bicrystal substrates by pulsed laser deposition.  Three ring samples were made from each YBCO film using standard optical photolithography techniques.  One ring was placed across a grain boundary so as to contain two grain boundaries (GB ring); two other rings were patterned on each of the two adjoining single crystals (grain rings).  All three rings have the same geometry with an outside diameter $2a$ of 3 mm, an inside diameter $2a_1$ of 2.8 mm giving a ring width $w$ of 0.10 mm, and a thickness $d$ of 200 nm.  The substrate was cut into three pieces, each containing just one ring.  Note that all three rings come from a single YBCO film and as a result, the films \emph{per se} should have the same properties, such as current density, pinning force, etc.  Bicrystal substrates with 1.8$^{\circ}$, 2.8$^{\circ}$, 5.1$^{\circ}$ (two samples), and 7$^{\circ}$ [001]-tilt boundaries were used to make GB and companion grain ring samples.  Sometimes the current density can be diminished by external degrading factors, such as cracks on a sample or by maltreatment.  To cross-check the deduced current density values of grain rings, some grain rings were made into an open circuit by etching a line across the 100 $\mu$m width (open rings).  This changes the geometry of the current path without changing the properties of the superconductor; consistent values of $J_c$ were obtained. 

Magnetic measurements were conducted with a SQUID-based Quantum Design MPMS-7 magnetometer.  An individual sample was mounted on a Si disk with Duco cement and placed in a Mylar tube for support.  For each ring, the magnetic moment was measured as a function of temperature and magnetic field.  For temperature sweep experiments, a magnetic field was applied parallel to the c-axes of YBCO film at 5 K (500 Oe, 3 kOe, and 3 kOe for grain, GB, and open rings, respectively); the field levels were chosen to ensure each sample geometry was fully penetrated by the field. Subsequently the applied field was reduced to zero ($H_{app}$ = 0) to induce circulating currents in the material (and the magnet was ``reset'' to provide the quietest and most stable magnetic environment).  Then we measured the remanent-state magnetic moment as a function of temperature from 5 K to 95 K in 1 K steps.  Complementary field-dependent moments $m(H)$, i.e., hysteresis loops, were measured in increasing and decreasing applied magnetic fields in the range from 0 Oe to 30 kOe.  The field sweep measurements were conducted at temperatures of 5, 10, 20, 40, 60, 77, and 85 K for each ring.  To obtain the critical current density $J_c$ from the measured magnetic moments, the critical state expressions Eq. 3-8 were used for each sample configuration.   

\section{RESULTS of THE TEMPERATURE SWEEP EXPERIMENTS}

For the temperature sweep measurements, each ring was prepared in the critical state by applying a large magnetic field and then reducing it to zero.  Two examples of these results for $J_c(T)$ in zero applied field are shown in Fig. 1, for a ring containing a 5.1$^{\circ}$ GB and its companion grain ring.  Both have the same $T_c$ near 93 K, as did all of the grain, GB, and open rings.  For the grain ring in Fig. 1, the current density $J_{c}^{Gr}$ was calculated using Eq.~\ref{eq:ring}.  These results are typical of those observed for the companion rings at 5 K in zero applied field, 34-40 MA/cm$^2$.  For the GB ring, the current density $J_{c}^{GB}$ was calculated using Eq.~\ref{eq:GBring}.  As evident in the figure, the values are strongly suppressed relative to those in the grain ring.  This is particularly so at low temperatures, where $J$ at 5 K in zero applied field lies near 1.6 MA/cm$^2$.  (The higher-$J$ GB data, shown as filled symbols, were obtained in finite decreasing magnetic field, as will be discussed below).  We reported previously \cite{Verebelyi01} that corresponding studies of 1.8$^{\circ}$ GB rings yielded current densities almost identical to those for the grain rings. It was argued that the similarity between the 1.8$^{\circ}$ GB and its companion grain ring is a consequence of the large numbers of twin boundaries in YBCO thin films.  

\begin{figure}[btp]
\begin{center}\leavevmode
\includegraphics[width=8.5cm]{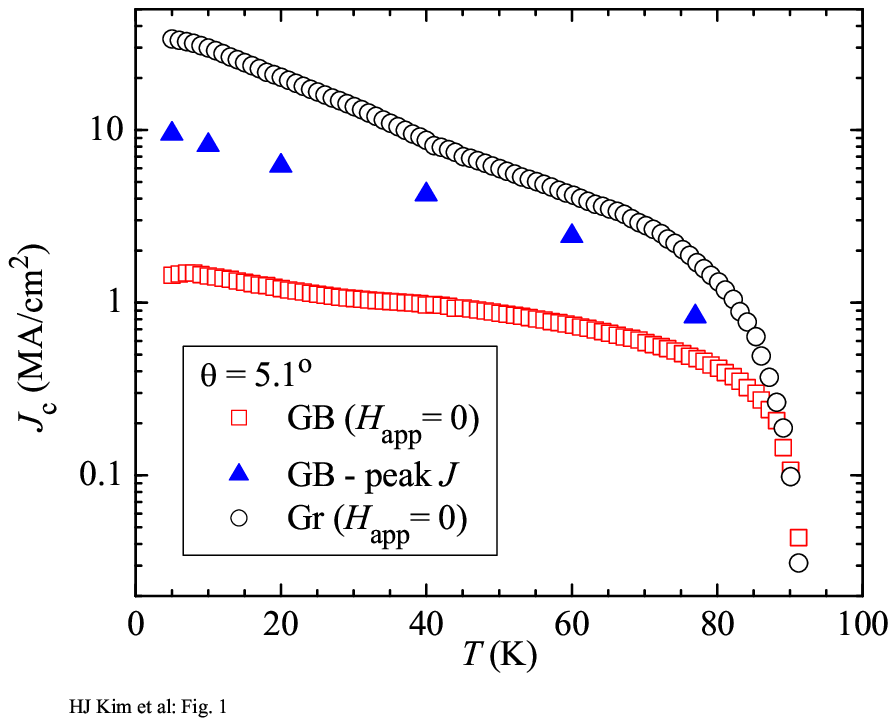}
\caption{ 
(color online) The critical current density $J_c$ of YBa$_2$Cu$_3$O$_7$ measured magnetically in zero applied magnetic field (open symbols), plotted versus temperature $T$.  Results are shown for YBCO rings on SrTiO$_3$ containing a single 5.1$^{\circ}$ [001] tilt grain boundary (GB) and for comparison, a grain ring (Gr) with no GB.  Filled symbols show peak values for $J_c$ in the GB ring measured in-field, as discussed in the text.
}
\label{figurename}\end{center}\end{figure}

As the misorientation of the grain boundary increases, the GB current density diminishes significantly.  While $J_{c}^{GB}$ is the same as the grain value for a 1.8$^{\circ}$ boundary, the values at $T$ = 5 K fall to $\approx$ 50 \% of $J_c^{Gr}$ at $\theta$ = 2.8$^{\circ}$, to 5 \% at 5.1$^{\circ}$, and to 3 \% at 7$^{\circ}$, respectively.  Interestingly (and fortunately for coated conductor applications), the \emph{fractional} transport is larger at $T$ = 77 K, rising to $\approx$ 60 \% of $J_c^{Gr}$ at $\theta$ = 2.8$^{\circ}$, to 15-20 \% at 5.1$^{\circ}$, and to 10 \% of $J_c^{Gr}$ at 7$^{\circ}$, respectively.  
These results show that a small angle c-axis tilt boundary in the range $1.8^{\circ} < \theta \le 5.1^{\circ}$ clearly impedes the current flowing across it.  More precisely, the range  $1.8^{\circ} < \theta \le  2.8^{\circ}$ contains the critical angle where a grain boundary begins to suppress the current flow across it. 

\section{RESULT OF THE FIELD SWEEP EXPERIMENTS}

The isothermal magnetic response was studied at temperatures $T$ = 5 - 85 K.  For the  grain and open rings, nicely symmetric curves of magnetic moment versus field were obtained, as illustrated by the inset to Fig. 2.  The symmetry of the hysteretic $m$ about the axis $m$ = 0 means that the same absolute magnitude 
of $J$ flows in the ring for increasing field (lower branch) and decreasing field (upper branch) histories. The critical current density was obtained 
from Eq.~\ref{eq:ring} and Eq.~\ref{eq:openring} for grain and open rings, respectively. Typical values were 34-40 MA/cm$^2$ in zero applied field at 5 K.  
Figure 2 shows the field and temperature dependence of $J_c^{Gr}$ for one of the grain rings. The dependence is simple, with a monotonic falloff with both $H$ and $T$.  As in temperature sweep experiments, the in-field features $m(H,T)$ for the 1.8$^{\circ}$ GB and its companion grain ring were nearly identical in field sweep experiments.  The curves of $m$ versus $H$ for the 1.8$^{\circ}$ and 2.8$^{\circ}$ GB rings were as symmetric as those of grain rings. 

\begin{figure}[btp]
\includegraphics[width=8.5cm]{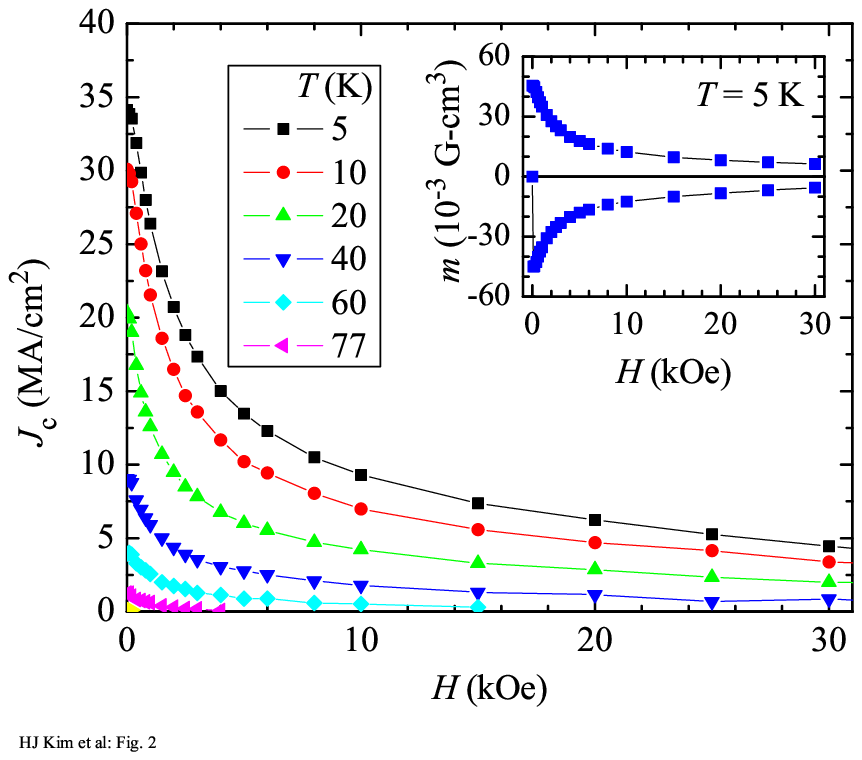}%
\caption{ 
(color online) For a YBCO grain ring without grain boundaries, the critical current density $J_c$ versus applied field $H$ at the temperatures shown.  Inset: the magnetic moment $m(H)$ at 5 K showing symmetric response in increasing and decreasing field histories.
}
\end{figure}

\begin{figure}[btp]
\includegraphics[width=8.5cm]{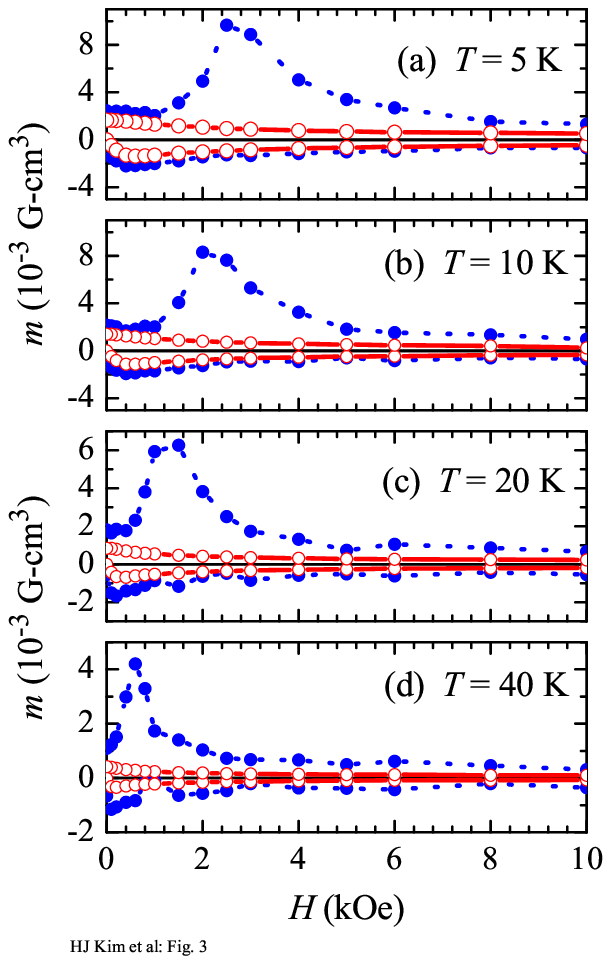}%
\caption{ 
(color online) The magnetic moment $m(H)$ for a GB ring on SrTiO$_3$ containing a single 5.1$^{\circ}$ GB.  The response is quite asymmetric, with a large peak in $m$ and $J_c^{GB}$ in decreasing field.  For comparison, the moment of an open ring (open symbols, heavy line) is included to show the approximate magnitude of the signal produced by ``U-turn'' grain currents.
}
\end{figure}

In contrast, the magnetic response of the rings with 5.1 and 7$^{\circ}$ grain boundaries is considerably more complex.  This is evident in Fig. 3, which shows $m(H)$ for the 7$^{\circ}$ ring at several temperatures.  Also included for comparison are virgin curves for a representative open ring, which shows the signal produced by a ring with a ``grain boundary'' of zero conductivity. Qualitatively, the ``excess'' magnetic moment in the GB ring (larger than the open ring) arises from currents crossing the GB and flowing around the circumference of the ring.   The most prominent feature in the 
$m(H)$ curves for the 5.1$^{\circ}$ and 7$^{\circ}$ GB rings is the appearance of a large peak in the decreasing field branch; by comparison, the loops for the open ring are symmetric about the $m$ = 0 axis, as this signal arises from grain-type currents. Compared with the GB current density in zero applied field, the value at the peak is considerably larger and it occurs at applied fields of several kOe magnitude.  To illustrate the difference, the values of $J_c^{GB}$ at the peak are included in Fig. 1 as solid symbols.  

The appearance of a peak in the GB current density has been reported many times and it is attributed to the effect of a magnetic field on a Josephson junction, which a grain boundary resembles.\cite{Evetts88}  The occurrence of a peak in the $m(H)$ curves of a GB sample, occurring only in decreasing field, marks the appearance of weak-link behavior\cite{Shantsev99}
in the small angle grain boundary.  It is well known that the maximum tunneling current flows across a Josephson junction when the net magnetic flux, which is perpendicular to the tunneling current flow, become zero on the area of the junction.  Applying this idea, one expects that the total magnetic field on a grain boundary $H_{local}$ becomes roughly zero at the peak in $J_c$.  Two major fields acting on the grain boundary are the applied magnetic field $H_{app}$ and the field $H_{self}$ created by induced currents flowing in the vicinity of and parallel to the grain boundary:  
\begin{equation}
H_{local}= H_{app} \pm |H_{self}|	
\label{eq:Hsum}				
\end{equation}
In increasing field, the directions of those two fields are the same, giving $H_{local}= H_{app}+|H_{self}|$.  On the other hand, decreasing the applied field reverses the currents near the grain boundary, giving $H_{local}= H_{app}-|H_{self}|$ and allowing a cancellation of the applied and self fields.  At the peak where $H_{local} \approx 0$, one has that $H_{peak} \equiv  H_{app, peak} \approx |H_{self}| \propto J$; the last proportionality follows from the fact that $H_{self}$ is created by currents flowing near the GB.

\begin{figure}[btp]
\includegraphics[width=8.5cm]{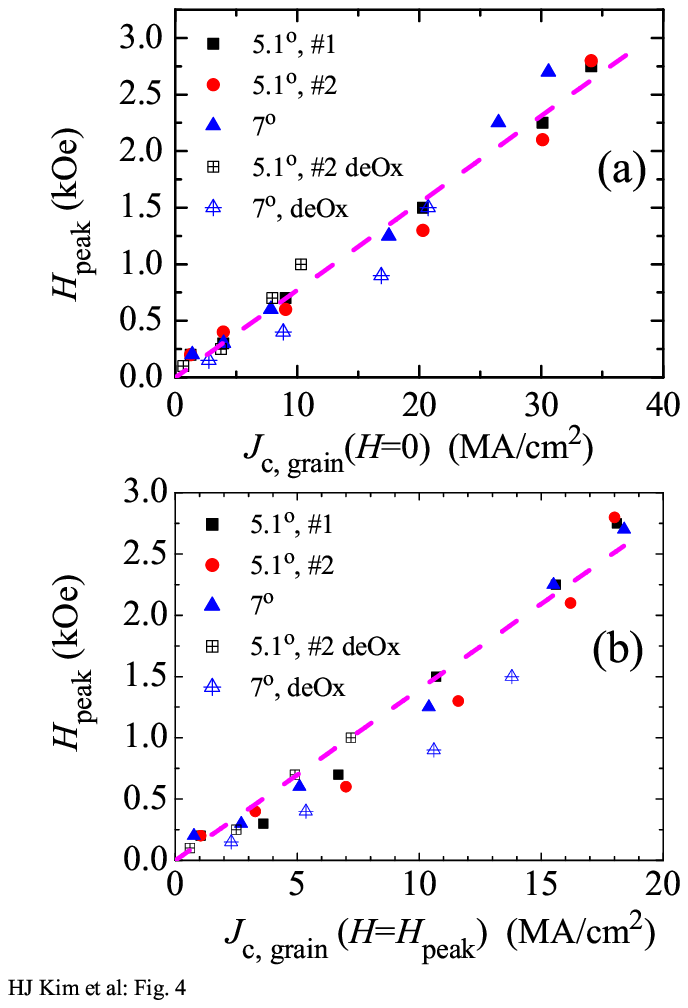}%
\caption{ 
(color online) The field $H_{peak}$ where the GB $J_c$ is a maximum in decreasing field, plotted versus \emph{grain} current density, for the samples shown at temperatures 5-60 K.  Grain boundary angles are [001] c-axis tilts. (a) The current density is $J_c^{Gr}$ measured in zero applied field, $H_{app}$ = 0 and (b) the current density is the grain value measured at the peak field, i.e., at $H_{app}$ = $H_{peak}$ for each case. 
}
\end{figure}

Figure 4 tests this scenario by plotting the peak field $H_{peak}$ as a function of grain current density $J_c^{Gr}$, where both are measured at temperatures from 5 K to 77 K.  In Fig. 4a, we consider the simplest  approximation that the effective currents have the density $J_c^{Gr}$ of the grains, measured in $H_{app}$ = 0.  This is the most appropriate choice, since grain currents near the GB (where $H_{local} \approx 0$) contribute most strongly to $H_{self}$.  Recall that the YBCO film is fully penetrated by flux and that $J_c^{Gr}$ is determined in separate measurements on companion rings.  Indeed, there is a strong correlation between the peak position $H_{peak}$ and the grain current density, as shown by the straight line in the figure with correlation coefficient $R^2$ = 0.983.  The figure includes data for the weakly linked GBs with $\theta$ = 5.1 and 7$^{\circ}$.  To test further the conjecture that a null local field produces the peak in $J$, we reduced the current density in two sets of rings while maintaining the same overall geometry;  they were partially deoxygenated by annealing in 0.2 bar of O$_2$ at 500 $^{\circ}$C, followed by furnace-cooling.  This decreased $T_c$ to $\approx$ 65 K for the 5.1$^{\circ}$ GB and 75 K for the 7$^{\circ}$ GB.  Data for those GBs are also included in Fig. 4 (labeled ``deOx'') and they follow the same trend as the fully oxygenated rings.

As seen in Fig. 4a, there is a clear correlation between $H_{peak}$ and the grain $J_c^{Gr}$.  One might consider, however, that the appropriate scale of nearby currents is the density $J_c^{Gr}$ measured at $H_{app}=H_{peak}$.  Thus Fig. 4b presents the peak field as a function of this lower, in-field current density.  The resulting plot is similar to that in Fig. 4a, with regression coefficient $R^2$ = 0.986. Overall, these analyses show that the position of the peaks in $J_c^{GB}$ tracks the nearby current density quite well.  This supports the conjecture that the peaks correspond to a nulling of the local field acting on a weakly linked grain boundary. 

Next we ask whether currents in the film can create local fields comparable with the observed $H_{peak}$.  This is a difficult question, as it involves the magnetic field $\emph{very}$ near the edge of a thin sheet of current-carrying superconductor. For one estimate, let us consider the case at $T$ = 5 K with grain current density $J_c^{Gr}$ $\approx$ 35 MA/cm$^2$, 
where $H_{peak} \approx$ 2.8 kOe (Fig. 4a).  
Numerical work of D\"{a}umling and Larbalestier \cite{Daeumling88} has shown that the perpendicular field is $\mu_0 H \approx 1.1 \mu_0 J_c d$ (in units of tesla) at the edge of a thin disk with radius/thickness = 10$^3$; this expression gives a field of $\sim$ 1 kOe at the edge of one disk and $H_{self} \approx$ 2 kOe, since current flows on both sides of the grain boundary.
An alternative estimate comes from the work of Brandt et al., who consider a very long thin strip of type II superconductor in a perpendicular magnetic field.\cite{Brandt93}  For the parameters cited, their expression for the normal field near the edge of a single strip gives values of 1.2 kOe at a distance $10 \times \xi_{ab}$ = 15 nm from the edge of the strip, and 1.6 kOe at distance $\xi_{ab}$.  Doubling these values as above to account for currents on each side of the grain boundary yields self fields $H_{self}$ that are very comparable with the 2.8 kOe observed experimentally.  While the geometries (disk, edge of thin strip) differ somewhat from the GB geometry, these estimates give some quantitative support for the ``null field-weak link'' model for the peak in GB current density.

\begin{figure}[btp]
\includegraphics[width=8.5cm]{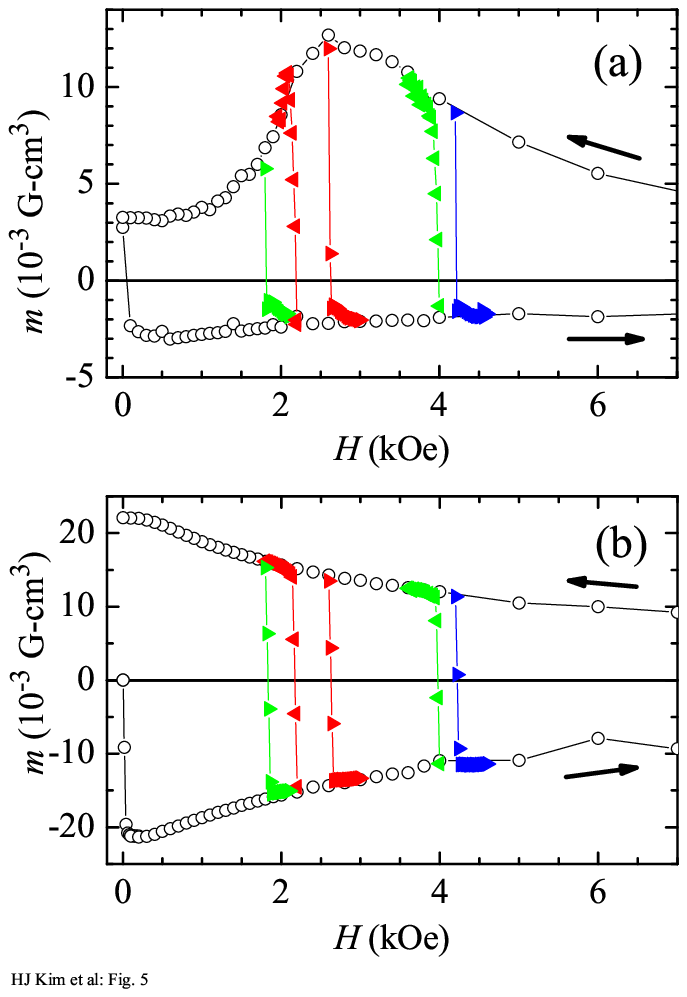}%
\caption{ 
(color online) Minor loops $m(H)$ at $T$ = 5 K for (a) a grain boundary (GB) ring with $\theta$ = 5.1$^{\circ}$ and (b) a grain ring with no grain boundary.  In both cases, the open symbols show the envelope signal obtained by monotonically sweeping the field from $H$ = 0 up to high field, then down to zero; filled symbols show the effect of 20 Oe steps in $H$ in the reverse direction, where the orientation of the symbols shows the direction of field change.  For the grain ring in (b), the response is symmetric, while it is quite asymmetric for the GB ring in (a).
}
\end{figure}

To probe further the grain boundary system near the peak in $J_c^{GB}$, we performed minor loop experiments on the 5.1$^{\circ}$ GB ring at 5 K.  The results are shown in Fig. 5a; for comparison, identical measurements on one of the grain rings are shown in Fig. 5b.  The following was done.  After making a standard $m(H)$ loop to 10 kOe and back to $H$ = 0, we began to retrace the loop, increasing $H$ from low field.  Then at 2.2 kOe, the applied field was decreased in 20 Oe steps until the magnetic moment reached the decreasing field branch of the $m(H)$ curve.  (The orientations of the symbols in Fig. 5 show the direction of field change.)  The changes in $m$ were very gradual and a field change of at least 100 Oe was needed in order to reach the upper branch.  This experiment in the increasing field branch was repeated at 4 kOe with similar results.  In contrast, the GB system is much more sensitive in the decreasing field branch: when decreasing $H$ from high fields to the peak at 2.6 kOe, an increase of 20 Oe produced a large reduction in the magnetic moment.  An increase of only 20 Oe is large enough to switch the magnetic moment almost to the lower branch. At decreasing fields of 1.8 and 4.2 kOe that lie on either side of the peak, the effect is similar although less dramatic at 1.8 kOe where $J_c^{GB}$ is smaller.  A smaller feature to note is the ``foot'' where a minor loop rejoins the main $m(H)$ curve; particularly noticeable at the lower branch, this component of $m$ develops as currents with density 
$J_c^{Gr}$ fully penetrate the ``strip-like'' portion of the ring.  

The major conclusion of the minor loop study is that near the peak in decreasing field, the grain boundaries are very sensitive to a reversal in field-sweep direction. Such a reversal induces oppositely directed perimeter currents in the film and switches the sign in 
Eq.~\ref{eq:Hsum}  from ($-$) to (+).  Those changes quickly add magnetic flux on the grain boundary, and this degrades the tunneling current across the GB.  In contrast, the same experiments on a grain ring give a symmetric response, as illustrated in Fig. 5b.  Steps in $H$ of 20 Oe give uniform changes in $m$, whether increasing or decreasing; the slope $dm/dH$ is the same as that at $H$ = 0, as is evident from the equally spaced points for $m$.  The latter is determined almost entirely by the geometry of the ring through its (effective) demagnetizing factor $D$. From this perspective, one can consider that the GB ring at its peak has the very large demagnetizing factor of a thin flat ring, but increasing $H$ in a minor loop tends to ``open up'' the GB and reduce $D$ to the smaller values more characteristic of a thin strip.  This discussion is, of course, only qualitative since some current does flow across the GB.  Overall, these minor loop experiments give further support to the picture that the asymmetric, history-dependent $J_c^{GB}$ and its peak in decreasing field all originate in the weak-linkage of a grain boundary that is strongly affected by the $\emph{local}$ field.

Some aspects of the magnetic field dependence remain difficult to understand.  In particular, analysis of the magnetic moments in Fig. 3 for increasing field history suggests that the values are about a factor-of-2 smaller than one might expect from estimates of 
$H_{local}$ combined with the observed behavior in the decreasing field branch.  This ``excess'' asymmetry was also observed by D\"{a}umling et al.\cite{Daeumling92} in transport studies on YBCO bicrystals with higher-angle grain boundaries. Other mechanisms, of course, contribute to the transport of critical currents across grain boundaries.  These can include pinning of GB vortices by facets on the boundary\cite{Cai98} or pinning of Abrikosov-Josephson vortices on the GB by interactions with (strongly pinned) Abrikosov vortices in the grains.\cite{Gurevich94,Kim00}.  However, none of these other mechanisms readily accounts for a strongly \emph{asymmetric} response, most notably the pronounced peak in the decreasing field branch \emph{only}.

\section{ESTABLISHING the CRITICAL STATE}

While the results cited so far were obtained with the entire sample fully penetrated with flux, it is informative to examine how the critical state is established.  For these experiments, the films were prepared in the virgin state by cooling them to low temperature in zero applied field.  Then the magnetic moment was measured as the field was increased in steps of a few Oe.  For the grain films patterned into narrow rings, the magnetic response is rather simple: $m$ increases almost linearly with $H$ with a slope $dm/dH \approx a^3$.  This is illustrated in Fig. 6a, where the initial slopes lie within 3 \% of the values calculated using Table I of Brandt \cite{Brandt97} with (inner radius $a_1$)/(outer radius $a$) = 1.4 mm/1.5 mm = 0.933.  This near-linearity is observed from $H$ = 0 up to the field of full penetration $H_p \approx 0.1 \times J_c d$ for this geometry.\cite{Brandt97}  Full penetration, which is marked by a sudden departure from the nominally linear response in Fig. 6a, occurs at progressively lower fields as $J_c$ deteriorates with increasing temperature. 

\begin{figure}[btp]
\includegraphics[width=8.5cm]{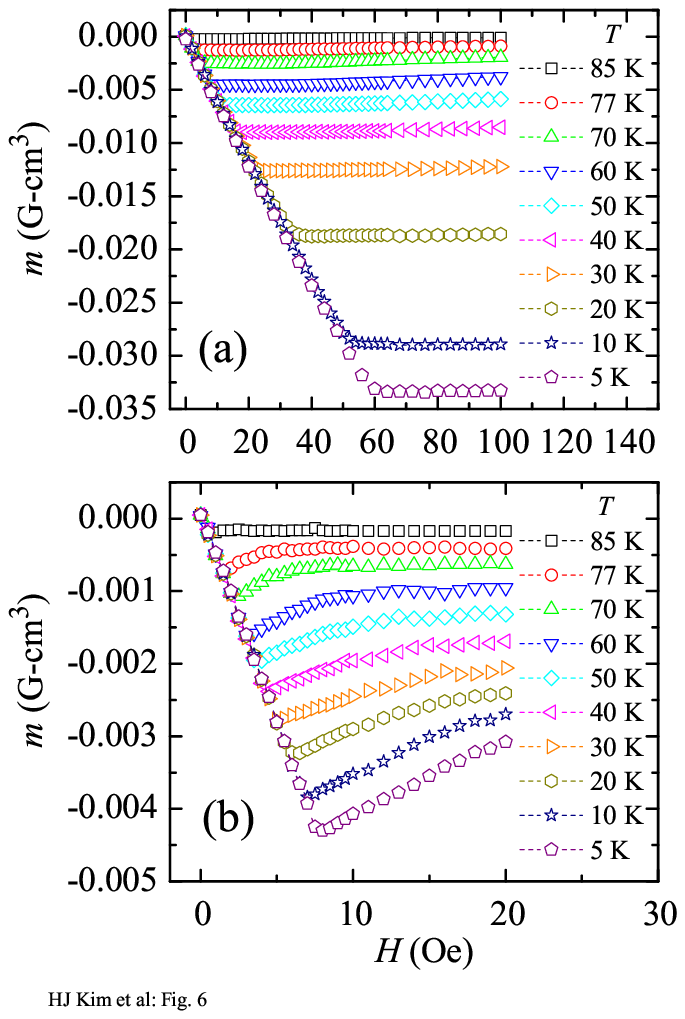}%
\caption{ 
(color online) The magnetic moment of virgin YBCO rings produced by applying small magnetic fields, at the temperatures shown.  (a) an epitaxial grain ring with no grain boundaries and (b) a ring with grain boundary angle $\theta$ = 5.1$^{\circ}$. 
}
\end{figure}

For rings containing a pair of small angle grain boundaries, the initial response is similar.  To show this, Fig. 6b presents $m(H)$ for a ring with $\theta$ = 5.1$^{\circ}$.  Compared with the grain ring, however, there are two qualitative differences.  First, the penetration field is much smaller, due to the lower $I_c$ of the GB; in the range shown, $H$ = 0 - 20 Oe, the film \emph{per se} is little penetrated by flux and the moment $m$ is determined almost entirely by circulating currents  that cross the GB.  Second, further increases in the applied field (above the GB penetration field) immediately reduce the magnitude of $m$.  Again, this can be attributed to the sensitivity of the GB $J_c$ to the \emph{local} field, i.e., applied plus self-field. Experimentally, the GB ring traps no flux until the applied field exceeds the (negative) peak in $m$.

Establishing the critical state in an open ring is an interesting contrast.  For a thin strip of width $w$ and length $\ell$ that the open ring resembles, the magnetic moment is given by \cite{Brandt97}
\begin{equation}
m(H) = -(J_c d w^2 \ell /4c) \tanh (H/H_c)
\label{eq:stripmH}					
\end{equation}
where the scaling field $H_c = J_c d / \pi $ and $J_c$ = constant.  Figure 7 shows the initial magnetization curve $m(H)$ for an open ring at 5 K where, in comparison to the closed ring, the initial slope is much smaller and the approach to saturation is much more gradual.  The solid line shows Eq.~\ref{eq:stripmH} drawn with the dimensions $w, d, \ell$ of the strip and the current density value $J_c$ = 35 MA/cm$^2$ obtained from a full hysteresis loop, giving $H_c$ = 22.5 kA/m = 283 Oe.  The agreement at low fields is excellent.  In larger fields, the theoretical curve lies somewhat below the experimental $m$ 
because $J_c$ is not constant, but rather decreases with $H$.  

\begin{figure}[btp]
\includegraphics[width=8.5cm]{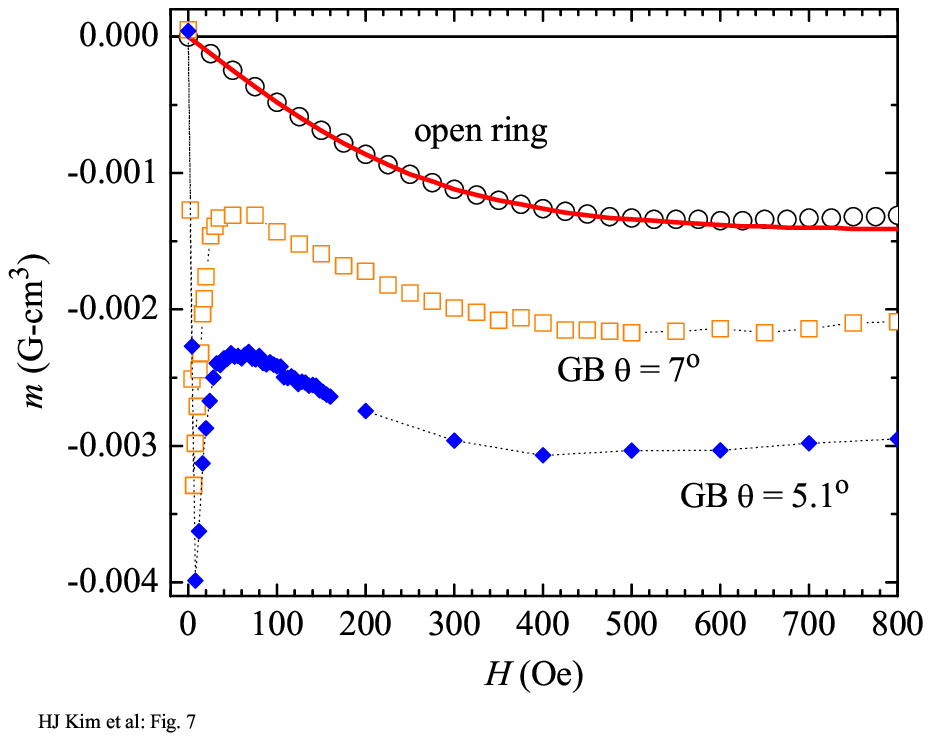}%
\caption{ 
(color online) The magnetic response $m(H)$ of an open ring; solid line shows a fit to Eq.~\ref{eq:stripmH} in the low field region where flux is penetrating the 'strip.'  Also included for comparison are the response of rings with $\theta$ = 5.1 and 7$^{\circ}$.  All data are at $T$ = 5 K.
}
\end{figure}

Also included in Fig. 7 are the magnetic responses of GB rings with $\theta$ = 5.1 and 7$^{\circ}$, all at 5 K.  The initial signal ($H$ = 0 - 10 Oe) develops very rapidly; this increase and the subsequent falloff of $|m|$ are the same as those shown in Fig. 6b and discussed above.  For larger fields $H  \gtrsim$ 50 Oe, the magnitude of $|m|$ again increases.  This additional contribution comes from currents induced and flowing entirely within the grain material (not crossing the GB), as evidenced by the fact that the $m(H)$ curves for the two GB rings are almost parallel to that for the open ring 'strip,' but displaced from it.  The same phenomenon - generating an additional contribution to $m$ by inducing grain currents that are reflected at the GBs - produces the 'foot' on the minor loops in Fig. 5a.

\section{CONCLUSIONS}
	
Using magntic methods, we have studied current conduction and weak linkage in low angle grain boundaries.  Materials investigated were ring-shaped YBCO films with or without [001]-tilt boundaries.  Relating the measured magnetic moments to the current configurations in GB, grain, and open ring samples, we obtained the grain boundary current density $J_{c}^{GB}$ from the ring-like contribution to the magnetic moment.  The temperature sweep experiments show that a small misorientation $1.8^{\circ} < \theta \le  7^{\circ}$ significantly reduces the grain boundary current density $J_{c}^{GB}$.  
However, no difference was observed between the 1.8$^{\circ}$ GB and its companion grain rings.  This study suggests that to obtain the highest current density in YBCO thin films and coated conductors with present grain boundary structures, it will be necessary to reduce the c-axis tilt grain boundaries into the range $\sim 3-4^{\circ}$.  
For the 5.1 and 7$^{\circ}$ GB rings, the $m(H)$ (and $J_c(H)$) curves have a large peak in finite field, but only for decreasing field history.  In small increasing field, the $m(H)$ curve of a GB ring resembles that of an open ring.  These two results from the field sweep experiments arise from the weak-linkage in moderately low-angle grain boundaries.  The high sensitivity of the current density across a GB to field changes on the GB strongly bolsters the weak-link interpretation of low-angle grain boundaries.

\begin{acknowledgments}
JRT wishes to acknowledge fruitful discussions with E.H. Brandt, A.A. Gapud, A. Gurevich, H.R. Kerchner, A. Palau, and A. Sanchez. We thank J. Budai for measurement of the grain boundary angles and M. Feldman for providing the 2.8$^{\circ}$ substrate.
This research was performed at the Oak Ridge National Laboratory, managed by UT-Battelle, LLC for the United States Department of Energy under contract No. DE-AC05-00OR22725.
\end{acknowledgments}


\section{References}

\bibliographystyle{prsty}

\end{document}